\documentclass[twoside]{procinasan}

\usepackage[utf8]{inputenc}
\usepackage{wrapfig}
\usepackage{changes}
\usepackage{color}
 
\newcommand{\nean}{\mbox{$\mathrm{{^{22}Ne}(\alpha,n){^{25}Mg}}$}}
\newcommand{\mac}{$\mathrm{\dot{M}_{acc}}$}
\newcommand{\mdon}{$\mathrm{\dot{M}_{don}}$}
\newcommand{\md}{$\mathrm{{M}_{don}}$}
\newcommand{\ma}{$\mathrm{{M}_{acc}}$}
\newcommand{\mwd}{\mbox{${\rm {M}_{\rm WD}}$}}
\newcommand{\msun}{\mbox{${\rm M_\odot}$}}
\newcommand{\isotope}[2]{${}^{#1}$#2}

\def\am{AM~CVn}
\input{procinasan.def}

\begin{document}

\pagebreak

\thispagestyle{titlehead}

\setcounter{footnote}{0}
\setcounter{equation}{0}
\setcounter{section}{0}
\setcounter{figure}{0}
\setcounter{table}{0}

\titlen{Thermonuclear oubursts of AM CVn stars}
{Yungelson~L.R$^1$, Piersanti~L.$^2$, Tornamb\'{e}~A.$^3$,
Cristallo~S.$^2$}
{$^1$Institute of Astronomy of the RAS, Moscow, Russia\\ 
lev.yungelson@gmail.com\\
$^2$INAF-Osservatorio Astronomico d'Abruzzo, Teramo, Italy\\ 
$^3$INAF-Osservatorio Astronomico di Roma, Monte Porzio Catone, Italy}

\abstre{We consider initial stage of the evolution of AM CVn type stars with white dwarf donors,
which is accompanied by thermonuclear explosions in the layer of accreted He. It is shown that the accretion
never results in detonation of He and 
accretor	s in \am\ stars finish their evolution as massive WDs. We found, for the first time,  that in the outbursts 
the synthesis of $n$-rich isotopes, initiated by the \nean\ reaction becomes possible.}

\selectlanguage{english}

\baselineskip 12pt
\section*{1. Introduction}
\begin{figure*}[t!]
\centering
\includegraphics[width=0.5\textwidth,angle=-90,clip =]{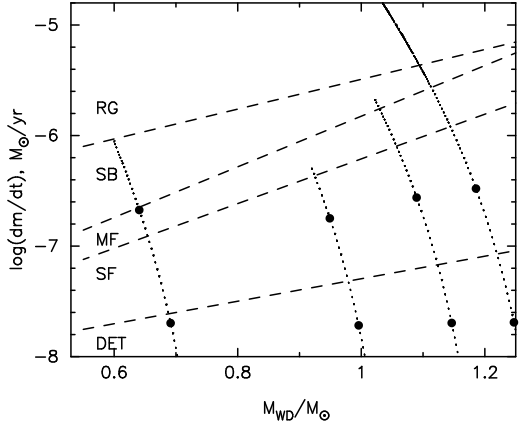}
\caption{Helium burning regimes in the layer of accreted He depending on the mass of the accretor 
\mwd\ and the accretion rate \mac\ \cite{2014MNRAS.445.3239P}.
Dashed lines mark the boundaries of the zones in which detonation (DET), strong (SF) and weak (MF) 
flashes, steady burning (SB) occur. In the RG-region of the diagram, extended envelopes around WDs
are formed. Dotted lines show the evolutionary tracks of the stars considered in this paper. 
Filled circles on the tracks correspond to the age of the system $10^5$ and $10^6$ years. \hfill}
\label{fig:regimes}
\end{figure*}

\am\ type stars are rare ($\simeq$60 objects \cite{2018A&A...620A.141R})
cataclysmic variable stars (CV) with helium white dwarf (WD) or low-mass helium
star donors (see, e.g., \cite{2001A&A...368..939N}). In the present paper we 
consider the systems of the former kind. 
Evolution of \am\ stars is
governed by the loss of angular momentum due to gravitational waves radiation
\cite{1967AcA....17..287P}. Immediately after formation, the rate of mass-loss
by the WD donors in the \am\ stars can reach $\sim10^{-5}$ \msun/yr; later it
decreases to $\sim10^{-11}$\msun/yr \cite{2001A&A...368..939N}. Depending on
the accretion rate onto the WD \mac, the burning of He in the accreted layer may be
steady, occur in quasi-cyclic mild or strong  thermonuclear flashes 
(eventually accompanied by matter ejection). 
Also He-detonation is possible, producing an explosive event of SN Ia proportion 
\cite{1982ApJ...253..798N}. 
An extended envelope 
may form around the WD, if \mac\ exceeds steady-state burning rate of He.
Figure~\ref{fig:regimes} shows the dependence of the burning regimes of He 
in the \mwd-\mac\ plane 
according to our calculations \cite{2014MNRAS.445.3239P}. Thermonuclear flashes
accompany the earliest stage of the evolution of \am\ stars.

We have investigated the flashes on the accreting WDs in the \am\ type CVs with
donors -- helium WDs and the accompanying
nucleosynthesis \cite{2015MNRAS.452.2897P,2019MNRAS.484..950P}. The stars with
the initial masses of components (\ma+\md)=\\(0.6+0.17), (0.92+0.15), (1.02+0.20), and
(1.02+0.30)\msun\ were considered. Such masses are characteristic for known
detached binary WDs, possible ancestors of the \am\ stars (see discussion in 
\cite{2015MNRAS.452.2897P}) and fit the model of the population of \am\
stars \cite{2001A&A...368..939N}.\footnote{Further, we will distinguish models
by the initial masses of the components.}

In \S2 we present the results of calculations, which are summarised in \S3.

\section*{2. Results of computations}

{\sl Evolution.} The algorithm of computations of \mdon\ is described in
\cite{2001A&A...368..939N}. The evolution of accretors was calculated using 1-D
FUNS evolutionary code \cite{2006NuPhA.777..311S}.
The  nuclear network included
about 700 isotopes from \isotope{1}{H} to \isotope{210} {Bi}, coupled by $\simeq$1000
reactions \cite{2016ApJ...833..181C}. Expansion of the WDs during the flashes
was limited by RLOF. It was assumed that the excess of the matter is lost from
the system with the specific angular momentum of the accretor. 

Accretion is accompanied by the release of gravitational energy, which
``counteracts'' the cooling of the WD. Heating of WD outer He-rich layer causes
ignition of He. Since the matter in the burning zone is degenerate, a
thermonuclear runaway occurs. Radiative transfer can not ensure the removal of
released energy and there arises a convective zone propagating both 
inward and outward with respect to the He-ignition point.
Convection removes the heat from the
He-burning zone, but also mixes into it ``fresh'' He, stimulating burning. This
process is accompanied by an increase of the radius of the WD, which leads to
the RLOF and ejection of the matter enriched by the burning products into ISM.
\begin{figure*}[t!]
\includegraphics[width=.48\textwidth,clip]{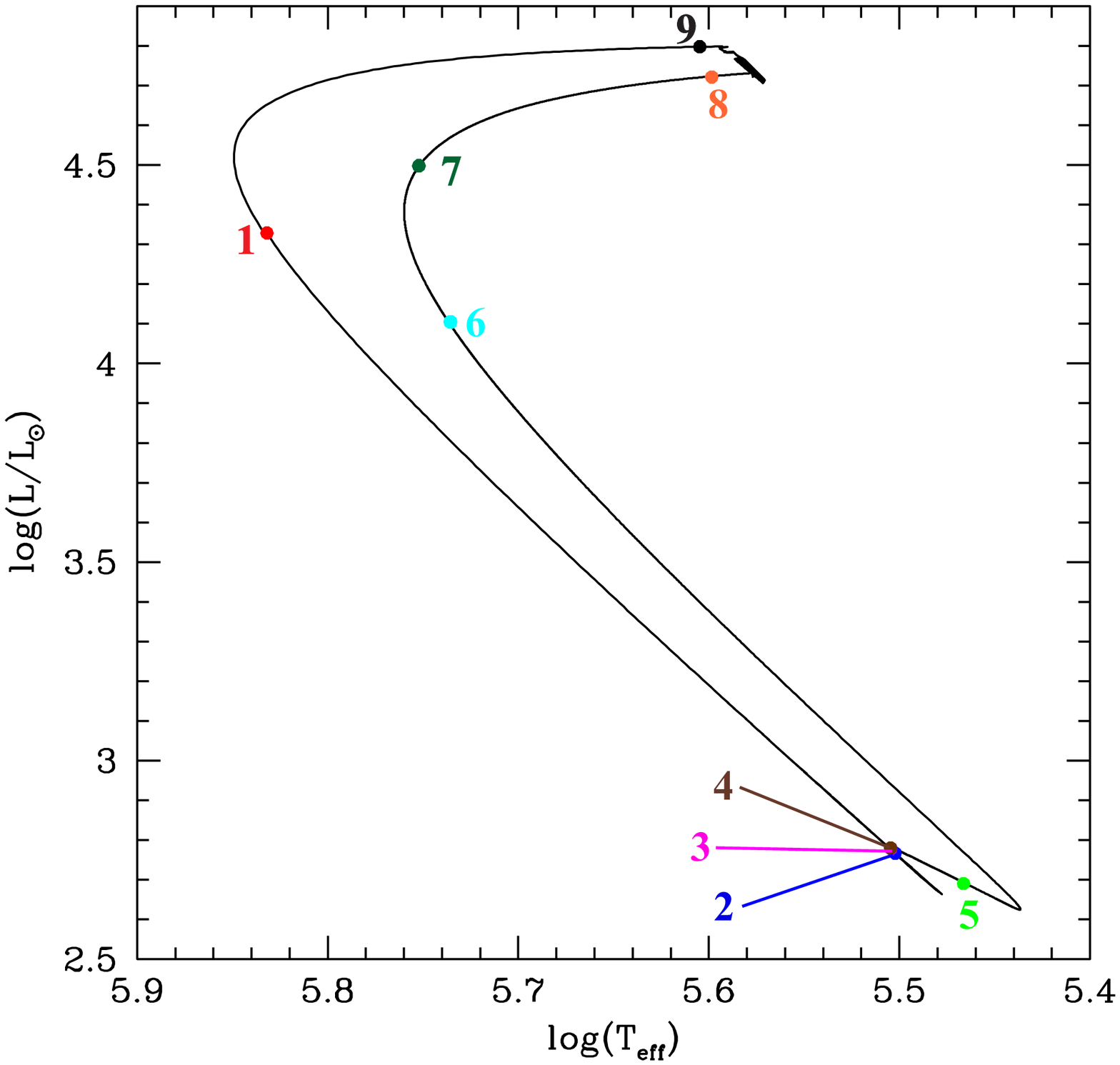}
\hskip 0.02\textwidth
\includegraphics[width=.48\textwidth,clip]{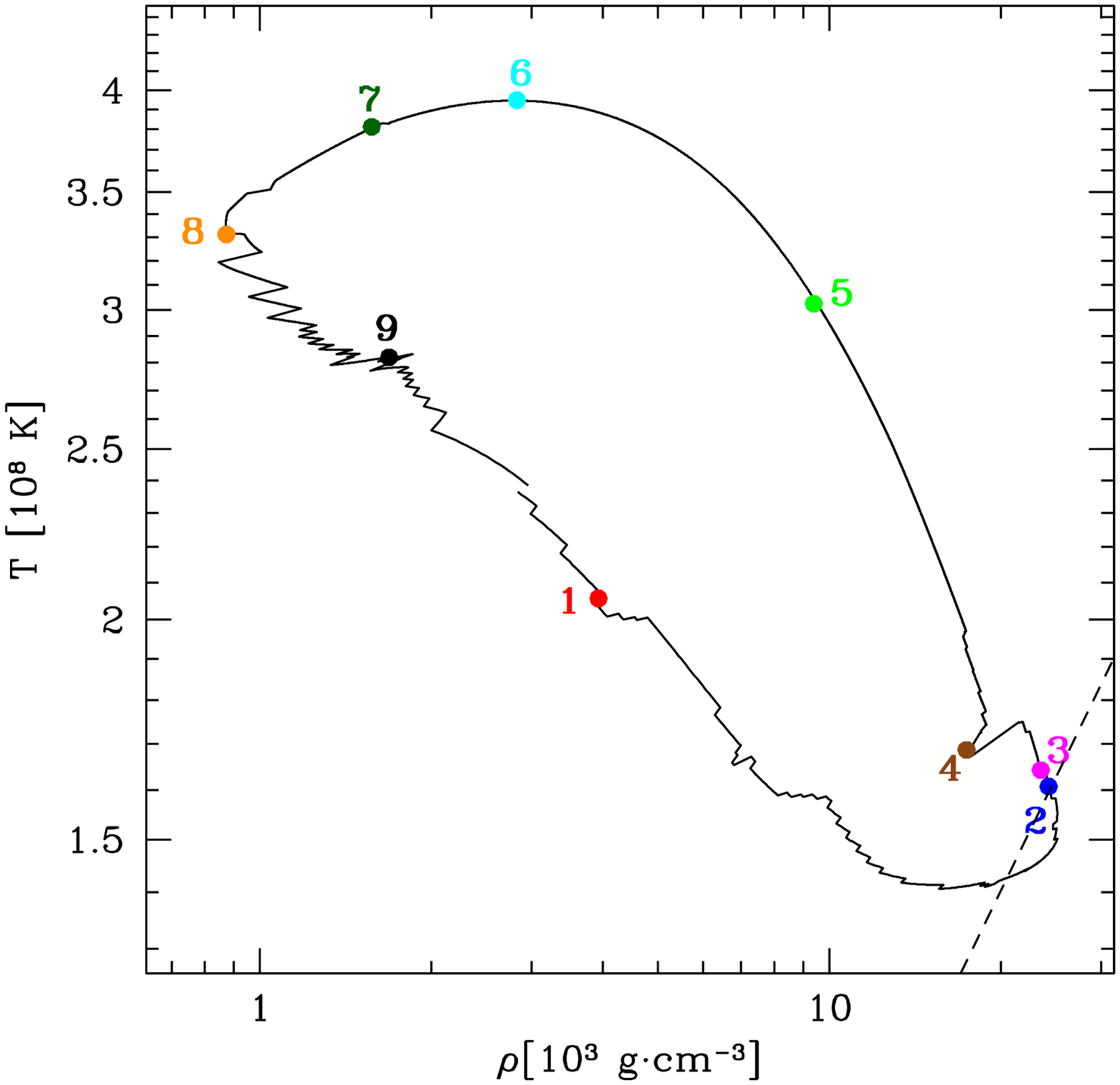}
\caption{Left panel -- evolutionary track of accreting WD in (1.02+0.3)~\msun\ system in the
HRD. Right panel -- evolution of temperature and density in the He-burning zone.
1 -- resumption of accretion after previous flash,  $t_{9-1}$=28 yr;
2 -- He-ignition, $t_{1-2}$=870 yr;
(2-3) -- rapid increase of nuclear energy release, $t_{2-3}$=5.93 yr; 
(3-4) -- thermonuclear runaway, formation of convective zone, $t_{3-4}$=214 day;
5 -- maximum outward extension of convective zone,
$T\approx3\cdot10^8$\,K, activation of \nean\ reaction, onset of neutron-isotopes nucleosynthesis, $t_{4-5}$=314 day;
6 -- temperature maximum in the He-burning zone ($3.8\cdot10^8$\,K), $t_{5-6}$=7.5 day;
(6-8) -- active burning of \isotope{22}{Ne}, $t_{6-8}$=3.57 yr; 
(8-9) -- RLOF stage, $t_{8-9}$=8.63 yr. \hfill}
\label{fig:loop}
\end{figure*}  

\begin{figure*}[ht!] 
\begin{minipage}[t]{0.5\textwidth}
\includegraphics[width=\textwidth]{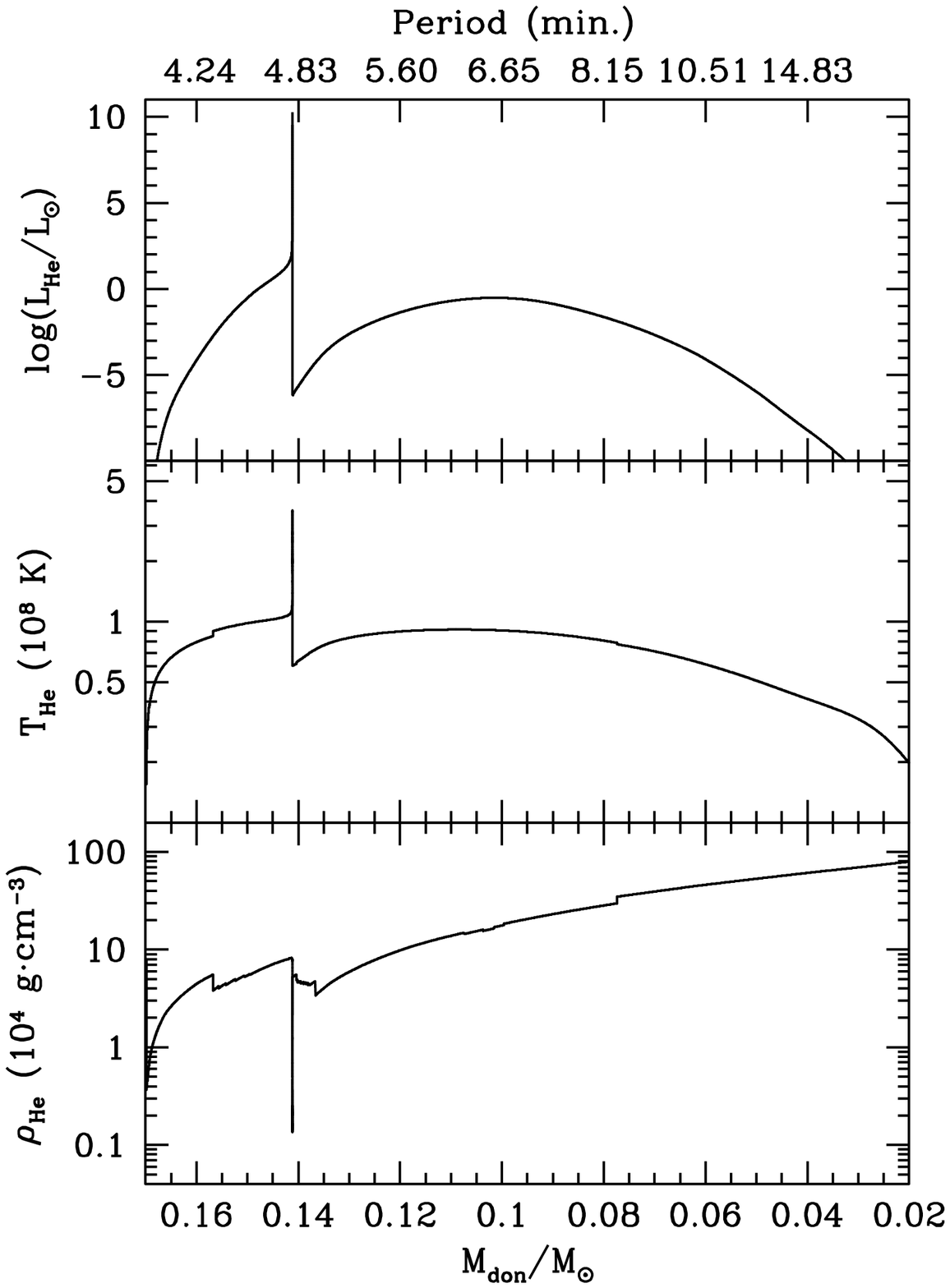}
\hskip 0.02\textwidth
\end{minipage}
\begin{minipage}[t]{0.5\textwidth}
\includegraphics[width=\textwidth]{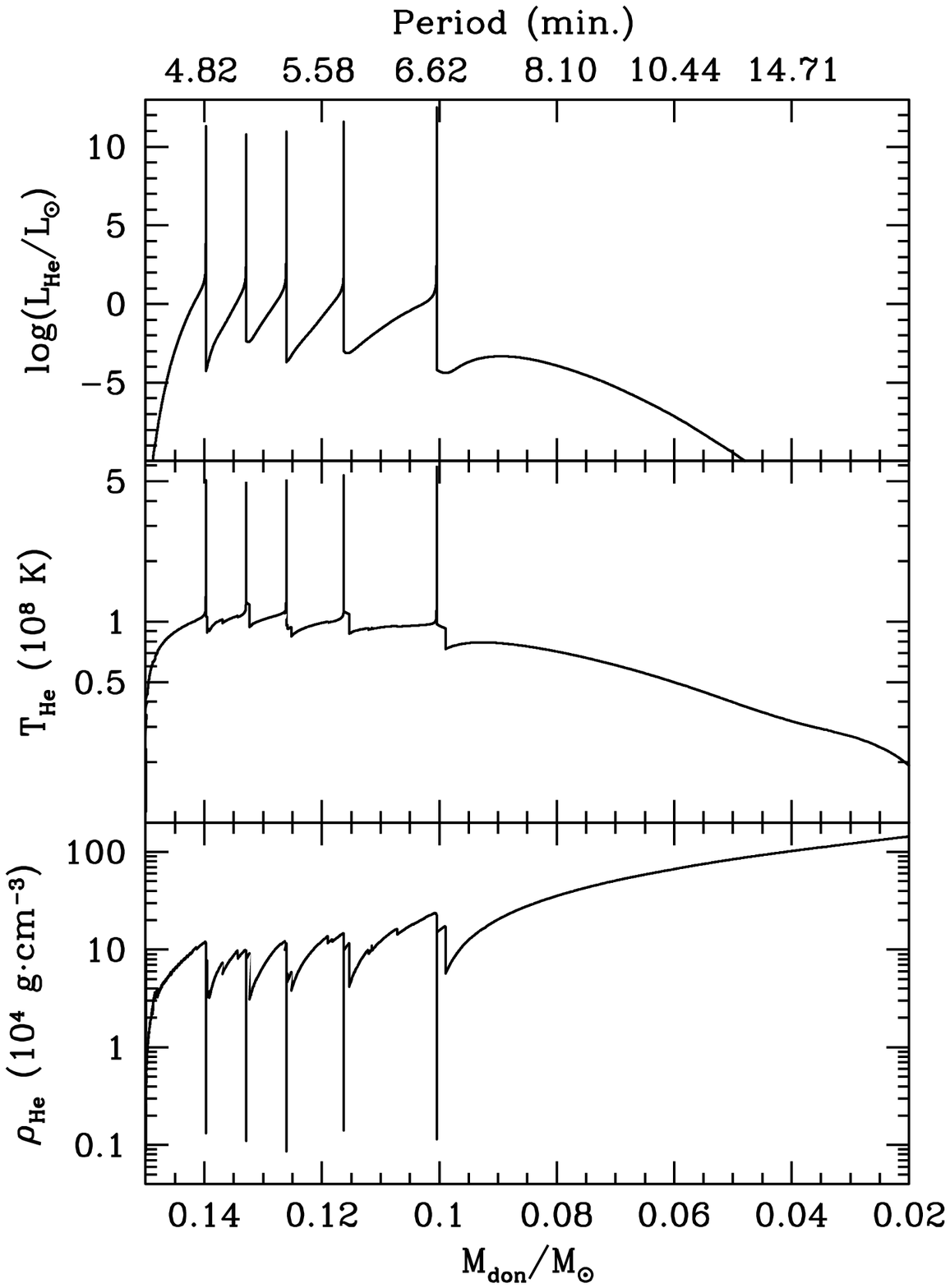}
\end{minipage}
\begin{minipage}[t]{0.5\textwidth}
\includegraphics[width=0.96\textwidth]{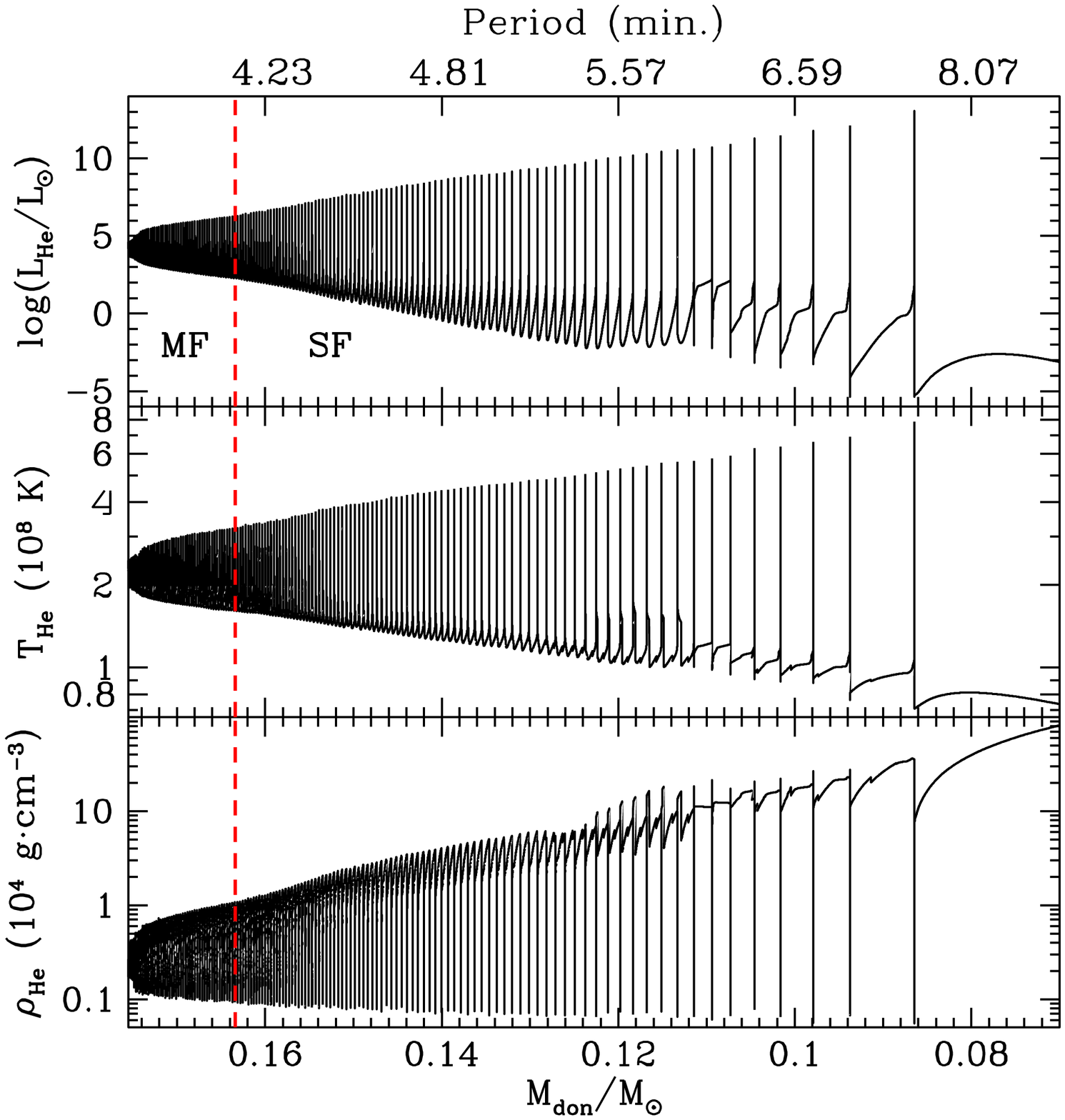}
\end{minipage}
\hskip 0.01\textwidth
\begin{minipage}[b]{0.49\textwidth}
\vskip -5.4cm
\hskip 10pt
\caption{Characteristics of the He-burning zone at the flashes stage as a function of 
donor mass for the 
systems (from left to right)  (0.62+0.17)\msun, (0.92+0.15)\msun, (1.02+0.30)\msun. 
Top to bottom: luminosity due to He-burning, temperature and density in the He-burning zone. 
For the system  (1.02+0.30)\msun\ vertical dashed line separates mild and strong flashes regimes.
The numbers along upper x-axis mark orbital periods of the system.} 
\end{minipage}
\label{fig:bursts}
\end{figure*}

Evolution of the dwarf during a typical cycle triggered by a 
strong flash is
shown in Fig.~\ref{fig:loop}. Figure~3 presents the evolution of the
characteristics of the burning layer for three selected models. In the system
(0.60+0.17)\msun\ one weak flash occurs  $5\cdot10^4$~yr after the start of
accretion and accumulation of 0.03\msun\ of He. After the flash, for
$\simeq2.5\cdot10 ^4$~yr the release of gravitational energy prevails over
cooling, but then, because of the fall of \mdon, the temperature in the He layer
starts to decrease, new flashes do not occur, and evolution ends by the formation of
a CO WD with a massive He envelope. In the system (0.92+0.15)\msun, the first
flash occurs  24000~yr after the start of accretion and accumulation of
0.01\msun\ of He. The flash turns out to be ``strong'', because \mac\ is low
(Fig.~\ref{fig:regimes}) and the degeneracy of  matter at the base of He layer 
by the time of the
flash is higher than in the other considered systems. Afterward, 4 more flashes occur. The time
interval between flashes and their power are increasing, because of the decline
of \mac. Evolution ends, like in the case of the system (0.60+0.17)\msun, by
transformation of the accretor into a massive CO WD with a thick He-mantle. In the system
(1.02 + 0.30)\msun, immediately after the contact \mac\ exceeds the rate of
stationary He-burning, but the matter excess is only $\simeq$0.01\msun\ and
the common envelope which can be formed hardly influences the  evolution of the system.
Progressively stronger flashes begin 120~yr after contact. Altogether, 50 mild and 87 flashes occur. 
Matter retention efficiency in flashes gradually 
decreases from 1 to 0. After the 87th strong flash, \ma$\approx1.1$\msun,
\md$\approx0.07$\msun. Further accretion does not lead to flashes, a massive
WD forms, like in the other cases.  

{\sl Nucleosynthesis.} During strong flashes the physical conditions 
are suitable for the activation of the {\it
intermediate neutron capture process (``i-process'')}, which occurs when the
number density of neutrons is $n_n\approx(10^{14}-10^{16})$\,cm$^{-3}$
\cite{1977ApJ...212..149C}. Until recently, {\it i}-process was associated with
the late AGB or post-AGB stars, still hypothetical ``rapidly accreting WDs''. As
a source of neutrons, the  \isotope{13}{C}($\alpha$,n)\isotope{16}{O} reaction was
considered, which occurs when  protons are ingested during a He-flash by convection 
into the hot He- and C-rich zone, resulting in the appearance of regions with high concentrations of
neutrons. Since in the \am\ stars WD accretes the matter that does not contain
hydrogen, \isotope{13}{C} is not synthesised. But as it is shown in 
Fig.~3,
already at the stage of mild flashes the temperature in
the He-burning zone attains $3\cdot10^8$\,K and in the case of strong flashes --
$10^9$\,K. Under such conditions, the \nean\ reaction is activated, which becomes
the source of neutrons for the {\it i}-process, resulting in the formation of
isotopes up to the stable \isotope{208}{Pb}.

\begin{figure*}
\begin{minipage}[t]{0.5\textwidth}
\includegraphics[width=0.95\textwidth]{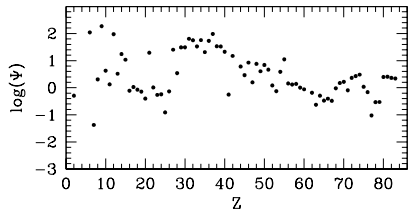}
\end{minipage}
\hskip 0.01\textwidth
\begin{minipage}[b]{0.49\textwidth}
\caption{The factor of elemental enrichment for the matter lost by the system 
(1.02+0.3)\msun\ in strong flashes.  \hfill}
\label{fig:ej}
\end{minipage}
\end{figure*}

During the flashes, WD loses all (or a part of) the layers with modified
chemical composition. The latter can be characterized by the factor of elemental enrichment
$\Psi=N_{el}/N_{ISM}$, where $N_{el}$ and $N_{ISM}$ are relative concentrations
of a certain element in the matter lost by the star and the ISM, respectively.
As an illustration, Fig.~\ref{fig:ej} shows the value of $\Psi$ for 
the matter lost by the system (1.02+0.30)\msun\ during strong flashes. 
In total, the
system lost 0.051\msun\ of the matter with chemical composition of accretor and
0.052\msun\ of the matter enriched in $\alpha$-elements and neutron-rich isotopes.

\section*{3. Conclusion}
\am\ stars with WD donors experience less than 100 flashes per system in the first $\sim10^5$~yr of existence.
Formation rate of such \am\ stars is $\sim10^{-3}$ per year
\cite{2001A&A...368..939N}. The flashes
occur predominantly on the most massive WDs, therefore, we can roughly estimate that 
currently in the Galaxy may exist
$\sim\,$10 \am\ stars experiencing thermonuclear flashes.

Even the most massive of the systems studied by us do not  experience the  ``last'', dynamic flash that destroys accretor.
Therefore, at least within the framework of our assumptions, the hypothetical optical transients 
SN\,.Ia \cite{2007ApJ...662L..95B} do not exist.

We found that the nuclear {\it i}-process triggered by the \nean\ reaction, which is usually associated with massive
stars, can occur during the outbursts of He burning on accreting WDs. However, rare  \am\
stars are vanishingly unimportant for the 
chemical evolution of the Galaxy. But their phenomenon shows that 
some exotic nuclear processes can occur in the depths of stars, hidden to the observers.

\small
\bibliography{yptc1}
\normalsize
\end{document}